\documentclass[12pt,letterpaper]{article}
\usepackage{amsmath,latexsym,amssymb,color,hyperref}
\newcommand{\be}{\begin{equation}}
\newcommand{\ee}{\end{equation}}
\newcommand{\bea}{\begin{eqnarray}}
\newcommand{\ena}{\end{eqnarray}}
\newcommand{\no}{\noindent}
\newcommand{\nb}{\nonumber}

\renewcommand\o{\omega}

\newcommand\m{\ensuremath{\mu}}

\newcommand\n{\ensuremath{\nu}}

\newcommand\tr{\text{tr}}

\newcommand{\de}{\partial}

\newcommand{\ba}{\begin{eqnarray}}
\newcommand{\ea}{\end{eqnarray}}

\makeatletter
\def\ps@mine{%
    \def\@oddfoot{\hfil\thepage\hfil}\let\@evenfoot\@oddfoot
    \let\@oddhead\@evenhead%
    \let\@mkboth\@gobbletwo
    \let\sectionmark\@gobble
    \let\subsectionmark\@gobble
    }
\renewcommand\section{\@startsection {section}{1}{\z@}%
                                   {-3.5ex \@plus -1ex \@minus -.2ex}%
                                   {2ex \@plus.2ex}%
                                   {\normalfont\large\sffamily\bfseries}}
\renewcommand\subsection{\@startsection {subsection}{1}{\z@}%
                                   {-3.5ex \@plus -1ex \@minus -.2ex}%
                                   {2ex \@plus.2ex}%
                                   {\normalfont\sffamily\bfseries}}
\makeatother

\begin{document}

\def\FILL{\hfill\hfill\hfill}

\title{\sffamily\bfseries 
FRW Cosmology in Ghost Free Massive\\ Gravity 
from Bigravity  \\[3ex]
  \normalsize
   D. Comelli$^a$, M. Crisostomi$^{b,c}$, F. Nesti$^{b,d}$ and   L. Pilo$^{b,c}$\FILL\\[2ex]
   \it\small
   $^a$INFN, Sezione di Ferrara,  I-35131 Ferrara, Italy\FILL\\
   $^b$Dipartimento di Fisica, Universit\`a di L'Aquila,  I-67010 L'Aquila, Italy\FILL\\
   $^c$ INFN, Laboratori Nazionali del Gran Sasso, I-67010 Assergi, Italy\FILL\\
   $^d$ ICTP, I-34100 Trieste, Italy\FILL\\[2ex]
   {\tt comelli@fe.infn.it}, {\tt marco.crisostomi@aquila.infn.it},\FILL\\
   {\tt fabrizio.nesti@aquila.infn.it}, {\tt luigi.pilo@aquila.infn.it}\FILL\\[-6ex]
}

\date{\small \today}

\maketitle

\thispagestyle{empty}

\vspace*{-6ex}

\def\abstractname{\sc Abstract}
\begin{abstract}
\no
  We study FRW homogeneous cosmological solutions in the bigravity
  extension of the recently found ghost-free massive gravity.
  When the additional extra metric, needed to generate the mass term, is taken as nondynamical and
  flat, no homogeneous flat FRW cosmology exists.  We show that, when the additional metric is a
  dynamical field a perfectly accetable FRW solution exists.  Solutions fall in two branches.  In
  the first branch the massive deformation is equivalent to an effective cosmological constant
  determined by the graviton mass.  The second branch is quite rich: we have FRW cosmology in the
  presence of a ``gravitational'' fluid.  The control parameter $\xi$ is the ratio of the two
  conformal factors. 
 When $\xi$ is small, the evolution is similar to GR and interestingly the universe flows at late
  time towards an attractor represented by a dS phase.
\end{abstract}

\bigskip

\section{Introduction}
Recently, there has been a renewed interest in the search of a modified theory of gravity at large
distances through a massive deformation of GR (see~\cite{Hint} for a recent review). A great deal of
effort was devoted to extend the seminal work of Fierz and Pauli (FP)~\cite{Fierz:1939ix} at
nonlinear level~\cite{Gabadadze:2010}.  The FP theory, defined at linearized level, is plagued by a
number of diseases. In particular, the modification of the Newtonian potential is not continuous
when the mass $m^2$ vanishes, giving a large correction (25\%) to the light deflection from the sun
that is experimentally excluded~\cite{DIS}. A possible way to circumvent the physical consequences
of the discontinuity was proposed in~\cite{Vainshtein}; the idea is that the linearized
approximation breaks down near a massive object like the sun and an improved perturbative expansion
must be used that leads to a continuous zero mass limit.  In addition, FP is problematic as an
effective theory. Regarding FP as a gauge theory where the gauge symmetry is broken by a explicit
mass term $m$, one would expect a cutoff $\Lambda_2 \sim m g^{-1} = (m M_{pl})^{1/2}$, however the
real cutoff is $\Lambda_5 = (m^4 M_{pl})^{1/5}$ or $\Lambda_3 = (m^2 M_{pl})^{1/3}$, much lower than
$\Lambda_2$ \cite{AGS}. A would-be Goldstone mode is responsible for the extreme $UV$ sensitivity of
the FP theory, that becomes totally unreliable in the absence of proper UV completion.  Recently it
was shown that there exists a non linear completion of the FP theory that is free of ghosts up to
the fourth order~\cite{Gabadadze:2011} and avoids the presence of the Boulware-Deser
instability~\cite{BD}.  Then the propagation of only five degrees of freedom and the absence of
instabilities was extended to the non perturbative level in~\cite{GF}; this was shown also in the St\"uckelberg
language in~\cite{deRham:2011qq}. The bigravity extension of this
thoery was shown to be ghost-free~\cite{HRBI}.

Quite naturally massive gravity leads to bigravity.  Indeed, any massive deformation, obtained by
adding to the Einstein-Hilbert action a non-derivative self-coupling of the metric $g$, requires the
introduction of an additional metric $\tilde g$. This auxiliary metric may be a fixed external
field, or be a dynamical one.  When $\tilde g$ is non dynamical we are dealing with \ae ther-like
theories; on the other hand if it is dynamical we enter in the realm of bigravity~\cite{DAM1},
originally introduced by Isham, Salam and Strathdee~\cite{Isham}.  The need for a second dynamical
metric also follows from rather general grounds. Indeed, it was shown in~\cite{DeJac} that in the
case of non singular static spherically symmetric geometry with the additional property that the two
metrics are diagonal in the same coordinate patch, a Killing horizon for $g$ must also be a Killing
horizon for $\tilde g$.  Thus, it seems that in order that the Vainshtein mechanism is effective and
GR is recovered in the near horizon region of a black hole, $\tilde g$ has to be
dynamical~\cite{ussphe}. In this paper we also show that cosmology calls for the bigravity
formulation of massive gravity. While in the St\"uckelberg formulation there is no homogeneous flat FRW
solution~\cite{FRWfroz}, (see also for a related work \cite{Alberte,Cham-Volk,cur,Koyama}) in the present
work we show that flat FRW homogeneous solutions do exist in the bigravity formulation.

In section \ref{bi} the formulation of massive gravity as bigravity is reviewed. The cosmological
ansatz is introduced in section \ref{eqcon}, where the structure of the modified Einstein equations
and the consequences of Bianchi identities are studied.  Cosmological evolution falls in two
branches described in section \ref{b1} and section \ref{b2}. In section \ref{curv}, the results of
the previous sections are extended to the case of spatially curved geometries.  Section~\ref{con}
contains our conclusions. The full set of Einstein equations can be found in the appendix.

\section{Massive Gravity and Bigravity}
\label{bi}

Any modification of GR that turns a massless graviton into a massive one calls for additional DoF
(degree of freedom). An elegant way to provide them is to work with the extra tensor $\tilde g_{\mu
  \nu}$. When coupled to the standard metric $g_{\mu \nu}$, it allows to build non-trivial
diff-invariant operators that lead to mass terms, when expanded around a background.  Consider the
action~\cite{DAM1}
\be
S=
\int d^4 x  \sqrt{\tilde g} \,   \kappa   \, M_{pl}^2\; \tilde {\cal R}+\sqrt{g} \left[ 
M_{pl}^2 \left( {\cal R}
-2  m^2   \, V \right)  + L_{\text{matt}} \right] ,
\label{act} 
\ee
where $R(g_i)$ are the corresponding Ricci scalars and the interaction potential $V$ is a scalar
function of the tensor $X^\mu_\nu = {g}^{\mu \alpha} {\tilde g}_{\alpha \nu}$.  Matter is minimally
coupled to $g$ and it is described by $L_{matt}$. The constant $\kappa$ controls the relative size
of the strength of gravitational interactions in the two sectors, while $m$ sets the scale of the
graviton mass. The action (\ref{act}) brings us into the realm of bigravity theories, whose study
started in the '60~\cite{Isham}.  An action of the form (\ref{act})
can be also viewed as the effective theories for the low lying Kaluza-Klein modes in brane world
models~\cite{DAM1}. The massive deformation is encoded in the non derivative coupling between
$g_{\mu\nu}$ and the extra tensor field $\tilde g_{\mu \nu}$. Clearly the action is invariant under
diffeomorphisms, which transform the two fields in the same way (diagonal diffs).  Taking the limit
$\kappa \to \infty$, the second metric decouples, and gets effectively frozen to a fixed background
value so that the ``relative'' diffeomorphisms are effectively broken, as far as the first metric is
concerned.  Depending on the background value of $\tilde g_{\mu \nu}$ one can explore both the
Lorentz-invariant (LI) and the Lorentz-breaking (LB) phases of massive gravity~\cite{PRLus}. When the second
metric is dynamical this is determined by its asymptotic properties, as discussed below. In this
case notice that $\tilde g_{\mu \nu}$ is determined by its equations of motion (for any finite
$\tilde M_{pl}$) so that we will be working always with consistent and dynamically determined
backgrounds.  The role played by $\tilde g_{\mu \nu}$ is very similar to the Higgs field, its
dynamical part restores gauge invariance and its background value determines the realization of the
residual symmetries.

The modified Einstein equations can be written as\footnote{When not
specified, indices of tensors related with $g$($\tilde g$) are raised/lowered with
$g(\tilde g)$}
\begin{gather}
\label{eqm1}
\,{E}^\mu_\nu  +  Q_1{}^\m_\n = \frac{1}{2\;M_{pl}^2 }\, {T}^\mu_\nu   \\
\label{eqm2}
\kappa  \, {\tilde E}^\mu_\nu  +  Q_2{}^\m_\n = 0 \; ;
\end{gather}
where we have defined $Q_1$ and $Q_2$ as effective energy-momentum tensors induced by the
interaction term. The only invariant tensor that can be written without derivatives out of $g$ and
$\tilde g$ is $X^\mu_\nu = g^{\mu \alpha}_1 {\tilde g}_{\alpha \nu}$ \cite{DAM1}. The ghost free
potential~\cite{Gabadadze:2011,HRBI}\footnote{A very similar potential having the same form but with $X$
  instead of $X^{1/2}$ was considered in~\cite{us}.}  $V$ is a
particular 
4-parameter scalar function of
$Y^\mu_\nu=(\sqrt{X})^\mu_\nu$ given by
%
\be
\label{eq:genpot}
\qquad V=\sum_{n=0}^4 \, a_n\, V_n \,,\qquad n=0\ldots4
\vspace*{-2ex}
\ee 
\vskip.3cm
\no
where The $V_n$ are the symmetric polynomials of $Y$
%
\be
\begin{split}
&V_0=1\,\qquad 
V_1=\tau_1\,,\qquad
V_2=\tau_1^2-\tau_2\,,\qquad
V_3=\tau_1^3-3\,\tau_1\,\tau_2+2\,\tau_3\,,\\[1ex]
&V_4=\tau_1^4-6\,\tau_1^2\,\tau_2+8\,\tau_1\,\tau_3+3\,\tau_2^2-6\, \tau_4\,,
\end{split}
\ee 
with $\tau_n=\tr(Y^n)$. In \cite{HRBI} it was shown that in the bimetric
formulation the potential $V$ is ghost free as in the St\"uckelberg formulation. We have that 
\bea
\label{eq:q1}
 Q_1{}_\nu^\mu &=&  { m^2}\, \left[ \;  V\; \delta^\mu_\nu \,  - \,  (V'\;Y)^\mu_\nu  \right]\\[1ex]
\label{eq:q2}
 Q_2{}_\nu^\mu &=&  m^2\, q^{-1/2} \, \; (V'\;Y)^\mu_\nu ,
\ena
where  $(V^\prime)^\mu_\nu = \de V / \de Y_\mu^\nu$ and  $q =\det
X=\det(\tilde g)/\det(g)$.

\section{Ansatz, Equations of Motions and Conservation Laws}
\label{eqcon}
For simplicity, here we consider the case of spatially flat geometries, the non flat case is
discussed in section \ref{curv}. 

In general we cannot set both metrics in diagonal form and preserve homogeneity. For instance, take
the following form for $g$ and $\tilde g$:
\be
\begin{split}
ds^2 &=   a^2(t) \left(- dt^2 +   dr^2 + r^2 \, d \Omega^2 \right) \\
\tilde{ds}^2 &= \omega^2(t) \left[- c^2(t) \, dt^2 +  2 D(t) \, dt\,
  dr + dr^2+  r^2 \, d \Omega^2 \right] \, 
\label{frw}
\end{split}
\ee
It is convenient to define
\be
\xi = \frac{\omega}{a} \; .
\ee
If $D \neq 0$, the metric $g$ is still homogeneous and the hypersurface of homogeneity is
$t=\text{const}$; on the other hand the metric $\tilde g$ is homogeneous with respect to a different
time slicing.  In addition, $Q_{1}$, $Q_{2}$ are not diagonal and ${Q_{1/2}}^r_r \neq
{Q_{1/2}}^\theta_\theta = {Q_{1/2}}^\varphi_\varphi$, as a result $Q_{1}$ and $Q_{2}$ are not energy
momentum tensors (EMT) for a homogenous perfect fluid. In addition, using Bianchi identities and the
equations of motion one can show that there is no physical solution when $D(t) \neq 0$. Indeed, from
the fact that the Einstein tensor for $g$ is diagonal, also $Q_1$ should be diagonal and with $D
\neq 0$ one finds that $\xi=\bar \xi $ is constant with $6 a_3 \, \bar \xi ^2+4 a_2 \bar \xi
+a_1=0$.\footnote{Notice that the constraint coincides with the one found for branch one solutions
  below, see sect.~\ref{b1}.}  Furthmore, Bianchi identities can be fulfilled only if $a_2 + 3 \,
\bar \xi \, a_3=0$.  Finally, in order to solve the tt component of Einstein equations for the
metric $\tilde g$ for any value of $r$, we are forced to set $D=0$.  Thus, we only need to consider
the case $D=0$.

The potential part of the action is separately invariant under diff and thus it gives two set of
Bianchi identities $\nabla_\nu {Q_1}^\nu_\mu=\tilde \nabla_\nu {Q_2}^\nu_\mu=0$ which are equivalent
to the following conservation law\footnote{One can easily show that diff invariance also implies
  that $\nabla_\nu {Q_1}^\nu_\mu=0 \Leftarrow \Rightarrow \tilde \nabla_\nu {Q_2}^\nu_\mu=0$.}
\be
\frac{d}{dt} \left( a^3 \, \rho_g \right) + p_g \, \frac{d}{dt}  a^3 =0
\, ,
\label{cons}
\ee
where
\be
\begin{split}
&\rho_g =\frac{m^2}{4 \pi  G} \left(3\, a_3 \,\xi^3+3 \,a_2\,
   \xi^2+\frac{3}{2}\, a_1 \,\xi +\frac{a_0}{2}\right) \, , \\
& p_g = -\frac{m^2}{4 \pi  G } \left[3\, a_3 \, c \, \xi^3+ a_2\, (2
   c+1) \,\xi^2+\frac{a_1}{2} \,(c+2) \, \xi+\frac{a_0}{2}  \right  ] \, .
\end{split}
\label{gfluid}
\ee
Eq. (\ref{cons}) is equivalent to the following condition 
\be
m^2\;\left(6 a_3 \, \xi^2+4 a_2 \, \xi +a_1\right) \left(c \, \omega  \,
  a'-a \, \omega'\right)=0 \, .
\label{bianchi}
\ee
The conservation of the matter EMT gives
\be
\frac{d}{dt} \left( a^3 \, \rho_m \right) + p_m \, \frac{d}{dt}  a^3 =0
\, .
\label{consm}
\ee
In order to determine the three functions of time $a, \;c,\;\o$ we need a set of three independent
equations. As in GR, one can show that for both metrics the time-time plus the space-space component
of the equations of motion leads to the Bianchi identities. As a result, also for massive gravity we
can take the conservation equations (\ref{bianchi}-\ref{consm}), together with
\bea
H^2 &&= \frac{8 \pi G}{3} \left (\rho_m + \rho_g \right) \, , \label{frw1}\\
\frac{{H_\omega}^2}{c^2} &&= \frac{m^2}{3 \kappa} \left(a_1 \,
  \xi^{-3}+6 \, a_2 \, \xi^{-2}  +18
  \, a_3   \, \xi^{-1} 
   +24 \, a_4 \right) \, ,
\label{frw2}
\ena
as independent equations. We defined the Hubble parameters in conformal time $H=a'/a^2$ and
$H_\omega= \omega'/\omega^2$.  The full set of equations of motion is given in Appendix \ref{meeq}.

Starting from eq.~(\ref{bianchi}) the cosmological evolution can be classified in two branches:
either there is an algebraic constraint for $\xi$
\be\
6 a_3 \, \xi ^2+4 a_2 \, \xi +a_1=0 \, ,
\label{alg}
\ee
or
\be
\frac{\o'}{\o}=c\;\frac{a'}{a} \, .
\label{bdiff}
\ee
Notice that, by definition $c(t)$ is the {\it positive} root of $c^2$ and thus $\omega'$ and $a'$
must have the same sign. Since in our universe $a'>0$, we will consider only the case $\omega'>0$.

\section{Cosmology: Branch one }
\label{b1}

Let us consider the case when the conservation equation (\ref{bianchi}) is solved as in~(\ref{alg}),
so that clearly $\xi(t)=\bar \xi$ is a constant satisfying the condition
\be 
6 a_3 \, {\bar \xi}^2+4 a_2 \, \bar \xi +a_1=0 
\label{cb1}
\ee
and $H_\omega = H \, {\bar \xi}^{-1}$.  

From (\ref{frw1}) we have
\be
H^2 =\frac{8 \pi G}{3} \, \rho_m + \frac{m^2}{3} \left[ a_0-6 {\bar \xi}^2
  \left(2 a_3 \bar \xi +a_2 \right)\right] \;. 
\label{tteq}
\ee
As a result, in this branch the effect of the massive deformation is to induce an effective
cosmological constant
\be
\Lambda_{eff} =\frac{ m^2 }{ 8 \pi G}  \left[ a_0-6 {\bar \xi}^2
  \left(2 a_3 \bar \xi +a_2 \right)\right] \; .
\ee
For what concerns the second metric, (\ref{frw2}) turns into an algebraic equation for $c(t)$ :
\be
c^2 = \frac{3 \kappa \, H^2 }{2 m^2 \left[6 \, \bar \xi  \left(2
      a_4 \, \bar \xi
   +a_3\right)+a_2\right]} \, .
\ee
Summarizing, branch one is equivalent to the usual FRW cosmology with a cosmological constant. This
is not very surprising.  Let us remind that in the bimetric
formulation, when one expands around flat space at the quadratic
level, there are two spin-2 modes: one is always massless by diff-invariance, the other has
Pauli-Fierz mass, see for instance~\cite{DAM1,PRLus}. The point here
is that condition~(\ref{cb1}) leads to a zero quadratic Pauli-Fierz mass.
The condition is the same found in~\cite{ussphe} in the branch where
no gravity modification are present. Notice that, even if at
the quadratic level the potential has no effect,  reappears in the theory 
at higher orders in perturbation theory.

Finally, let us comment on the $\kappa \to \infty$ limit which corresponds to freezing the second
metric.  From eq. (\ref{frw2}) one gets that $H_\o = 0$, but in the branch we are dealing with,
$\xi$ is constant and so also $H=0$. As result, for the branch one no flat FRW cosmology exists
when the second metric is non dynamical.

\section{Cosmology: Branch two }
\label{b2}

The Bianchi identities (\ref{bdiff}) in such a branch give
\be
c = \frac{H_\o}{H} \, \xi  \, .
\label{bnew}
\ee 
Inserting  the expression of $c$ in (\ref{frw2}) we have
\be
m^2 \left[6 \, \xi \left(4 \, a_4 \, \xi^2+3 \, a_3 \, \xi
   +a_2\right)+a_1\right]-3 \, \kappa \, \xi  \, H^2 = 0 \, .
\label{H2}
\ee
If we solve this equation for $H^2$ and compare with (\ref{frw1}), we find
\ba\nonumber
 &&m^2 \left[\xi ^2 \left(\frac{8\; a_4}{\kappa }-2\; a_2\right)+\xi 
   \left(\frac{6 \; a_3}{\kappa }-a_1\right)+\frac{a_1}{3 \; \kappa \, 
   \xi }+\frac{2 \; a_2}{\kappa }-2\;  a_3 \, \xi ^3-\frac{a_0}{3}\right] \\
&&=\frac{8 \pi  G \; \rho_m }{3}  \, .
\label{conII}
\ea
For matter with an equation of state $p_m=w \rho_m$, we have that $\rho_m=\rho_0 \,
a^{-3(1+w)}$. Thus, eq.~(\ref{conII}) can be solved for $\xi$ in terms of the conformal scale factor
$a(t)$.  The problem of the cosmological evolution is reduced to FRW cosmology in the presence of
matter plus an additional form of ``gravitational'' matter.  Though the solution of (\ref{conII})
can be written in a closed form, it is not particularly illuminating.

Alternatively, from the definition of $\xi$ and from (\ref{bnew}) we have
\be
\xi' = (c-1) \, a\,\, H \; \xi,
\ee
and from (\ref{conII}) we can express $a$ in terms of $\xi$: $a = f_a(\xi,w)$.  Finally, from
(\ref{eq1ss}) and (\ref{eq2ss}) (see appendix \ref{meeq}) $c$ can be expressed in terms of $\xi$
\ba
c\equiv 1+(1+w) \, f_c(\xi) \, ,
\ea
$f_c(\xi)$ is a function whose explicit form is not needed.  At this point, the velocity field of
$\xi$ is function of $\xi$ only
\be
\xi'=(1+w)\;f_c(\xi)\; H(\xi,w)\;f_a(\xi,w) \, \xi  \, ,
\ee
and it has two fixed points 
 \be
\begin{cases}
\xi=\bar \xi \, , \quad  c=1 \, \text{with }f_c(\bar \xi)=0 & \rho_m
\to 0 \\[.3cm]
\xi=\bar \xi \, ,  \quad \text{with } H(\bar \xi,w)=0 & \text{generic
  } \rho_m \end{cases} \, .
 \ee
 While the fixed point with zero Hubble parameter is not cosmologically very significant, the $c=1$
 case instead corresponds to a De Sitter phase.

 Far from the fixed points the main features can be deduced by a qualitative analysis of the
 solutions of (\ref{conII}). The variable $\xi$ is dimensionless and naturally is a function of $G
 \, \rho_m /m^2$.  In the early time, when $G \, \rho_m /m^2\gg 1$ we have two possible regimes:
 small $\xi\propto {\cal O}(m^2/G\,\rho_m)$ and large $\xi\propto {\cal O}(G\,\rho_m/m^2)$.  
 \vskip .8cm

\centerline{\bf Large $\mathbf{\xi}$}
\vskip .5cm
\no
The regime of large $\xi$ is determined by the leading positive power of $\xi$ present in
eq.~(\ref{conII}).  For simplicity, here we report only the leading results, the corrections to them
are ${\cal O}\left( m^2/ G\;\rho_m \right)$.  For $a_3\neq0$ the Leading term is $\xi^3$ and a
solution exists only if $a_3<0$, with
\ba
\xi&=&\left(\frac{8 \pi G\,\rho_m}{6 |a_3| m^2}\right)^{1/3}+\frac{4\;a_4-\kappa\;a_2}{3\;a_3\;\kappa} \, ;\\[1ex]
\rho_m+\rho_g&=&  2\,
\frac{a_4}{\kappa}\left( \frac{6 \;  m^2\; \rho_
    m^2}{\pi\, G\; a_3^2} \right)^{1/3} \, \ll \rho_m;\\[1ex]
 w_{eff}&\equiv&\frac{p_g+p_m}{\rho_g+\rho_m}=\frac{2\;w-1}{3}  
\, ;\\[1ex]
c &=& - w \, .
\ea
When $a_3=0$, the leading term is $\xi^2$ and a solution exists if $4\,a_4-k\,a_2>0$, with
\ba
\xi&=& \left[\frac{ 8 \pi \, G\, \kappa\,\rho_m}{
  m^2 \, 6(4 \,a_4- a_2 \,\kappa)} \right]^{1/2}\, ;\\[1ex]
\rho_m+\rho_g&=&  \frac{4 a_4}{(4\, a_4- \kappa \,a_2)}\rho_m \, ;\\[1ex]
 w_{eff}&=& w \, ;\\[1ex]
c&=& - \frac{3 w+1}{2} \, .
\ea
Finally, if $a_3=0$  and  $\kappa
\,a_2-4\,a_4=0$, the leading term is $\xi$ and  a large $\xi$
solutions exists when  $a_1<0$, with
\ba
\xi&=&-\frac{8\,\pi\, G}{3\; a_1 \;m^2}\;\rho_m \, ;\\[1ex]
\rho_m+\rho_g&=&  \frac{16 \,\pi\, G \;a_2}{3\; a_1^2\;m^2}\;\rho_m^2
\,\gg \rho_m ;\\[1ex]
w_{eff}&=& 2\; w+1 \, ;\\[1ex]
c&=& - (3 w+2) \, .
\ea
The large $\xi$ regime is characterized by non standard cosmology and/or negative values of $c$.
Except for the case where the quadratic term is dominating, the effective equation of state
$w_{eff}$ largely deviates from $w$, but even in this case $c<0$ when $w>0$. As a result, we
conclude that the large $\xi$ regime is not a physical one.

\pagebreak[3]

\vskip .8cm

\centerline{\bf Small $\mathbf{\xi}$}
\nopagebreak
\vskip .5cm
\nopagebreak
\no
Solutions with small $\xi$  are present only when $a_1> 0$  and in this case
\ba
\xi&=&\frac{a_1\; m^2}{8\,\pi\, G \, \kappa\,\rho_m} \, ;\\
\rho_m+\rho_g&=&  \rho_m \, ;\\ 
 w_{eff}&=&  w \, , \\
c&=& 4 + 3 w  \, .
\ea
For simplicity, we give only the leading terms, the corrections are ${\cal O} ( m^2/G\;\rho_m )$ and
can be sistematically computed.
In this regime cosmology is standard: once the matter is so diluted that $\rho_m$ is negligible in
(\ref{conII}), the system falls in the fixed point region and $\xi$ is almost constant; the universe
enters in a late time dS phase.

One can solve perturbatively (\ref{frw1}) to find $a$. For a radiation dominated universe we
find\footnote{As an example, in the following expressions we give the leading and the next to leading
  correction to $a$.}
\be
a(t)=\frac{t}{t_0}  +\frac{a_0 \, m^2 \, t^5}{30 \, t_0^3}+ \frac{m^4
  \, t^9 \left(a_0^2 \kappa +20 a_1^2\right)}{1080 \, \kappa \, 
    t_0^5} + \cdots \, , \qquad t_0
  = \left(\frac{3}{8 \pi G \, \rho_{0}}\right)^{1/2}.
\ee
In the case of matter dominated universe
\be
a(t)=\frac{t^2}{t_0^2} +\frac{a_0 \, m^2 \, t^8}{84
   \, t_0^6}+ \frac{m^4 \, t^{14} \left(4 \, a_0^2 \, \kappa +49 \,
     a_1^2\right)}{30576 \, 
   \kappa  \, t_0^{10}}+ \cdots   \, \qquad  t_0
  = \left(\frac{3}{2 \pi G \, \rho_{0}}\right)^{1/2}\,.
\ee 
\vskip.3cm
\no For instance, taking $\rho_0$ of order of the critical density today and $m \sim 10^{-33} eV$ we
have that during the radion era $\xi \sim 10 \, (1+z)^{-4}<<1$.

Let us also discuss the $\kappa \to \infty$ limit  in this branch. When the second metric is non
dynamical, from eq.(\ref{H2}) we get that $H \to 0$. Thus, also in this branch no flat FRW cosmology
exists with just a single dynamical metric.

\section{FRW with Spatial Curvature}
\label{curv}

Let us discuss the case with spatial curvature (see \cite{cur} for the case with frozen second
metric). The metrics take the form
\be
\begin{split}
ds^2 &=   a^2(t) \left(- dt^2 +   \frac{dr^2}{(1-k_1 \, r^2)} + r^2 \, d \Omega^2 \right) \\
\tilde{ds}^2 &= \omega^2(t) \left[- c^2(t)\, dt^2  + \frac{dr^2}{(1-k_2 \, r^2)}+  r^2 \, d \Omega^2 \right] .
\label{frwc}
\end{split}
\ee 
In the presence of curvature the conservation law for $Q_1$ and $Q_2$ take a different form
\bea
\left( F-1 \right) \, \left[ 2 \, c \, \xi \, \left(3 \, a_3 \, \xi +a_2
  \right)+2 \, a_2 \, \xi +a_1 \right] &=&0\,  ,
\label{bc1} \\[.3cm]
\nonumber
F(r) \, \xi \, \left \{ \left[2 \, \xi  \left(9 \, a_3 \, \xi +4 \, a_2\right)+a_1\right]
H_{\omega }-4 \, c \, H \left(3 a_3 \, \xi +a_2\right) \right \}-&&\\
c \, H \, \left(6 \, a_3 \, \xi ^2+8 \, a_2 \, \xi +3\,  a_1\right)+2
\,  \xi  \left(2 \, 
   a_2\,  \xi +a_1\right) H_{\omega } 
 &=&0\,,
\label{bc2}
\ea
where
\be
 F(r) =
\left(\frac{k_1 \, r^2-1}{k_2 \, r^2-1}\right)^{1/2} \, .
\ee
When $k_1 \neq k_2$, $F$ is different from one and the Bianchi identities must hold for any value of
$r$; this gives three nontrivial relations. From (\ref{bc1}), we can solve for $c$
\be
c= -\frac{2 \, a_2 \, \xi +a_1}{6 \, a_3 \, \xi ^2+2 \, a_2 \, \xi } \, ;
\ee
then inserting $c$ in
(\ref{bc2}) we get that
\bea
&&H_\omega =-\frac{2 \, H \left(2 \, a_2 \, \xi +a_1 \right)}{\xi  \left(18
    \, a_3 \, 
   \xi^2+8 \, a_2 \, \xi +a_1\right)}\, ;\\[.3cm]
&& H \, \left(2 \, a_2 \, \xi +a_1\right) \left(6 \, a_3 \, \xi ^2+4
  \, a_2 \, \xi +a_1\right){}^2 =0 \, .
\ea
The solution $2 \, a_2 \, \bar \xi +a_1=0$ would lead to $c=0$ and to
a degenerate second metric.  Thus, either $H=H_\omega =0$ or $\xi =
\bar \xi =const$, with
\be
6 \, a_3 \, \bar \xi^2+4 \, a_2 \, \bar \xi +a_1 = 0 \, .
\label{we}
\ee
In the special case $a_2=a_3=0$ we get the nonphysical solution $H=H_\omega=0$.
Thus, the only interesting solutions of Bianchi identities, when $k_1 \neq k_2$, are
the ones with $\xi= \bar \xi $  satisfying eq(\ref{we}). Notice  that for $\xi=\bar\xi$ we
have that $c=1$. The further constraint coming from the equations of motion gives
\be
\frac{k_2-k_1}{a^2} + \frac{8 \pi G}{3} \rho_m + \Lambda =0 \, ,
\ee
where $\Lambda$ is a function of $\bar \xi$. The previous equation is algebraic in $a$ and it gives
$a=$constant and $H=0$.  Thus, in order to obtain a reasonable cosmology we are forced to set $k_1 =
k_2=k_c$ and we are back to the Bianchi identity (\ref{bianchi}). The analysis of the previous
sections is the same, except for the presence of the curvature term.

Let us now discuss the differences with the St\"uckelberg approach  where $\tilde
g$ is non dynamical.  Formally, the non-dynamical limit would correspond to $\kappa \to \infty$.  To make
contact with the existing literature, we take $\tilde g$ equivalent to the Minkowski flat metric
implying clearly that $c$ and $\omega$ cannot be arbitrary. Imposing that the Riemann
curvature tensor of $\tilde g$ vanishes we find the conditions 
\be
k_c <0 \,,\qquad c= \frac{\o \, H_\o \,}{\sqrt{-k_c}} \, .    
\label{cflat}
\ee
Therefore, FRW cosmology with frozen second metric exists only with a negative non-vanishing spatial
curvature~\cite{FRWfroz}.  When (\ref{cflat}) holds, Bianchi identities can be realized only within
branch one, leading to eq.~(\ref{cb1}). 

\section{Conclusions}
\label{con}
In this paper we studied FRW cosmological solutions of the bigravity
extension of the recently found ghost free massive gravity
theory. It was shown in ref.~\cite{FRWfroz,cur} that when the auxiliary metric needed for the
formulation of the theory is taken to be non dynamical, no flat FRW cosmology exists.  On the other hand,
as we showed here, flat FRW solutions are allowed when the second metric is promoted to a dynamical field.  This
result, together with the analysis of spherically symmetric solutions~\cite{ussphe}, indicate that
bigravity is more than just a tool; it is an important ingredient in the formulations of a
physically acceptable theory of massive gravity.  The cosmological evolution is governed by Bianchi
identities (BI). Depending on the way they are realized, cosmology falls in two separate branches.
When the BI are implemented by an algebraic constraint on the ratio $\xi$ of the two scale factors,
we find standard FRW cosmology, in the presence of a cosmological constant proportional to the
graviton mass scale.  When BI are implemented by a differential relation between the Hubble
parameters of the two metrics, cosmology is instead richer. Cosmological evolution at the early
times, where $G\rho_m/m^2\gg1$, is controlled by the parameter $\xi$.  In the presence of matter
with an equation of state $w$, the large $\xi$ regime leads to
unphysical solutions with $c<0$.  
On the contrary,
the small $\xi$ regime is safe and reproduces the standard early FRW cosmology; at late time the
universe sets in a dS attractor region. The model is interesting and though there are a number of free
parameters in the potential, it is surprisingly predictive.  The next step is to study perturbations
to check whether massive gravity cosmology is viable.  We will report these results in a future work.

\vskip .7cm 
\centerline{\large \bf Note added} 
\vskip .3cm 
During the completion of this work, a study on the same topic appeared~\cite{vol}. Our results seem
to agree with~\cite{vol}, except for some exotic case.  Almost at the
same time of our paper, another paper~\cite{Hcosm} on the same subject was announced in the
arXiv.

\begin{appendix}

\section{Modified Einstein Equations}
\label{meeq}

For completeness we collect the full set of equations and we restore the presence of the spatial
curvature $k$.  The non vanishing equations for the metric $g$ are the $tt$ component and the
space-space components and read
\bea
&&m^2\left[6 a_3 \xi ^3 +6 a_2 \xi ^2 +3 a_1
   \xi  +a_0 \right]+8 \pi  G \rho _m =\frac{3 k}{a^2}+3 H^2  \, ;
\label{eq1tt}
\\[.3cm]
&&m^2 \left[6 a_3 c  \xi ^3+2 a_2 (2 c+1)  \xi ^2+a_1
   (c+2)  \xi +a_0 \right]-8 \pi  G p_m \nb\\
&&=\frac{k}{a^2}+\frac{2 H'}{a} + 3 H^2 . \label{eq1ss}
\ea
For the metric $\tilde g$ the structure is the same and we have
\bea
&& m^2 \left(\frac{a_1 }{\xi ^3}+\frac{6
   a_2 }{\xi ^2}+\frac{18 a_3 }{\xi }+24 a_4 \right) = \frac{3 \kappa  \, k}{a^2 \xi ^2}+\frac{3
   \kappa  H_{\omega }^2}{c^2} \, ; \label{eq2tt}\\[.3cm]
&&m^2 \left(\frac{a_1}{c \, \xi ^3}+\frac{4 a_2 }{c \, \xi
   ^2}+\frac{6 a_3 }{c \, \xi }+\frac{2 a_2 }{\xi
   ^2}+\frac{12 a_3 }{\xi }+24 a_4\right) \nb \\
&&= \frac{\kappa  \, k}{a^2 \, \xi ^2}+\frac{2 \kappa  \, H_\omega'}{a
  c^2 \, \xi }+\frac{3 \kappa \,  
   H_{\omega }^2}{c^2} -\frac{2 \kappa  \, c' \, H_{\omega }}{a \, c^3
   \, \xi}\, . 
\label{eq2ss}
\ea
Except for section \ref{curv}, in order to de-clutter formulae we have not included $k$; as we can
see from the previous equations the curvature can be easily restored without effecting the analysis.
A linear combination of eq. (\ref{eq1tt}) and its time derivative together with (\ref{eq1ss}) gives
(\ref{bianchi}) and the conservation law for matter. The same is true for (\ref{eq2tt}) and
(\ref{eq2ss}).
\end{appendix}

\end{document}